\documentclass[
  journal=pasa,
  manuscript=research-paper, 
  year=2020,
  volume=37,
]{cup-journal}

\usepackage{microtype,siunitx,booktabs}

\usepackage{amsmath,amsfonts,amssymb}	
\usepackage{subfig,color}
\usepackage{appendix}
\usepackage{xcolor}
\definecolor{menucolor}{rgb}{0.1,0.52,0.47}
\definecolor{urlcolor}{rgb}{0.85,0.37,0.01}
\definecolor{runcolor}{rgb}{0.46,0.44,0.701}
\definecolor{filecolor}{rgb}{0.2,0.5,0.01}
\definecolor{linkcolor}{rgb}{0.12,0.47,0.70}
\definecolor{citecolor}{rgb}{0.55,0.36,0.01}
\definecolor{anchorcolor}{rgb}{0.4,0.4,0.4}
\usepackage[colorlinks=true, urlcolor=urlcolor, linkcolor=linkcolor,citecolor=citecolor,filecolor=filecolor, anchorcolor=anchorcolor,menucolor=menucolor]{hyperref}
\definecolor{Grph1}{HTML}{67000D}
\definecolor{Grph2}{HTML}{000000}
\definecolor{Grph3}{HTML}{08519C}

\definecolor{Grp1}{HTML}{C51B7D}
\definecolor{Grp2}{HTML}{8C510A}
\definecolor{Grp3}{HTML}{00441B}
\definecolor{Grp4}{HTML}{7F2704}
\definecolor{Grp5}{HTML}{3F007D}
\definecolor{Grp6}{HTML}{08306B}
\definecolor{Grp7}{HTML}{67000D}
\definecolor{Grp8}{HTML}{525252}

\title{Multi-layered characterization of hot stellar systems with  confidence}

\author{Souradeep Chattopadhyay}
\affiliation{Department of Statistics, Iowa State University, Ames, IA 50011, USA.}
\alsoaffiliation{Department of Mechanical Engineering, Iowa State University, Ames, IA 50011, USA.}

\author{Steven D. Kawaler}
\affiliation{Department of Physics and Astronomy, Iowa State University, Ames, IA 50011, USA.}

\author{Ranjan Maitra}
\affiliation{Department of Statistics, Iowa State University, Ames, IA 50011, USA.}
\email[Ranjan Maitra]{maitra@iastate.edu}

\doi{10.1017/pasa.2020.32}

\received {dd Mmm YYYY}
\revised  {dd Mmm YYYY}
\accepted {dd Mmm YYYY}
\published{22 September 2020}

\keywords{methods: statistical -- astronomical instrumentation,
  methods, and techniques; methods: data analysis -- astronomical
  instrumentation, methods, and techniques; galaxy: globular clusters:
  general -- the galaxy}

\begin{document}

\begin{abstract}
	Understanding the physical and evolutionary properties of Hot Stellar
	Systems (HSS) is a major challenge in astronomy. We studied the dataset on 13456 HSS of
	\citet{misgeldandhilker11} that includes 12763 candidate globular
	clusters using stellar mass ($M_s$), effective radius ($R_e$)
        and mass-to-luminosity ratio ($M_s/L_\nu$), and found multi-layered homogeneous grouping among these stellar systems. Our methods elicited eight homogeneous ellipsoidal groups at the finest
	sub-group level. Some of these groups have high overlap and were merged
	through a multi-phased syncytial algorithm motivated from
	\citet{riveraandmaitra20}. Five groups were merged in the first phase,
	resulting in three complex-structured groups. Our algorithm determined
	further complex structure and permitted another merging phase,
	revealing two complex-structured groups at the highest level.
	A nonparametric bootstrap procedure was also used to estimate the
	confidence of each of our group assignments. These assignments
	generally had high confidence in classification, indicating great
	degree of certainty of the HSS assignments into our complex-structured
	groups. The physical and kinematic properties of
	the two groups were assessed in terms of $M_s$, $R_e$, surface
        density and $M_s/L_\nu$. The first group consisted 
	of older, smaller and less bright HSS while the second group consisted
	of brighter and younger HSS. Our analysis provides novel insight
	into the physical and evolutionary properties of HSS and also helps 
	understand physical and evolutionary properties of candidate globular
	clusters.
	Further, the candidate globular clusters (GCs) are seen to have very
        high chance of really being GCs rather than dwarfs or dwarf
        ellipticals that are also indicated to be quite distinct from
        each other.  
\end{abstract}


\section{Introduction}

Over the past many decades, astronomers have identified
commonalities in clusters of stars ranging from small groups to
large galaxies.  Hot stellar systems (HSS) are a class of
celestial objects consisting  of globular clusters, nuclear
star clusters, compact ellpitical galaxies, giant elliptical
galaxies, ultra compact dwarf elliptical galaxies, nuclear
star clusters and so on. These HSS are very important to
understand processes such as the formation of stars or
black holes, evolution of galaxies and so on.  
Indeed, the physical properties of these objects have been
directly linked to galaxy interactions and have been
extensively studied. One of the most useful
concepts~\citep{bursteinetal97, bernardietal03,kormendyetal09, misgeldandhilker11} in studying these objects is  the set of fundamental 
plane relations~\citep{brosche73}. These planes are
typically constructed with parameters such as luminosity,
surface brightness, stellar magnitude or central velocity
dispersion and help understand important
properties of these stellar 
systems. Different HSS subgroups typically have different fundamental
planes, photometric properties and other types of
interpretations~\citep{kormendy85, djorgovski95, mclaughlinetal05,
	meylanetal01, forbesetal08, harrisetal95, webbink85, ichikawa86,
	kormendyetal09, benderetal92}. But the origins of these relations,
the underlying evolutionary processes involved and the statistical
reliability of these results are not well understood and remain as
significant challenges. A huge volume of data has been collected and
catalogued in the last fifteen years to effectively address these challenges. For example  \citet{jordanetal08} catalogued and studied 12763
candidate globular clusters (GCs) from the Virgo Cluster
Survey during the eleventh Hubble Space Telescope observation
cycle. A more comprehensive catalogue was compiled by \citet{misgeldandhilker11} which included 693 additional HSS along with those of \citet{jordanetal08}. \citet{chattopadhyayandkarmakar13} pointed out that several studies
have previously compared stellar systems
such as globular clusters and dwarf spheroidals using two-point correlations  between different
projections of the fundamental plane of galaxies. 
Given our lack of understanding about HSS groups, their evolution
and relationships, an important step forward is a firm understanding of
which groups are demonstrably different in observational parameter spaces,
and which may be more closely related. These important questions can be
answered by modern multivariate statistical analysis methods, and one
effort to do so was by \citet{chattopadhyayandkarmakar13} who used  
$k$-means clustering with the jump
statistic~\citep{sugarandjames03} and found five and four
homogeneous groups in the 693 non-candidate and the larger 13456 HSS datasets.

Clustering~\citep{kettenring06, xuandwunsch09, everitt11}  is
an  unsupervised learning technique that groups observations
without a response variable. While there are many kinds of
clustering algorithms, most of them can broadly be categorized into
hierarchical and non-hierarchical approaches. Hierarchical
clustering algorithms yield a tree-like grouping hierarchy
while non-hierarchical algorithms, such as from $k$-means or model-based
clustering (MBC) methods, typically optimize an objective
function using iterative greedy algorithms -- these algorithms
typically require a specified number of groups. The objective
function is often multimodal and requires careful
initialization \citep{maitra09}. A detailed review on
MBC is provided, for instance, in
\citet{mclachlanandpeel00} or \citet{melnykovandmaitra10};
additionally, 
\citet{chattopadhyayandmaitra17} reviewed it in the context of  astronomical applications. Even though model-based methods are an
improvement over traditional hierarchical and non-hierarchical methods,
they are still mostly limited to finding regular-structured groups. For
example, the most common technique of Gaussian-mixtures 
MBC (GMMBC) assumes spherical or ellipsoidally-structured
spreads in their groups. Very often, however, the underlying groups in
a dataset are irregular in shape and do not all neatly fit the assumptions
underlying MBC. Thus advanced methods are necessary to identify
and analyze complex-structured groups. \citet{riveraandmaitra20}
proposed a novel method of identifying such groups by using {\em 
syncytial clustering} which incorporates the results from standard 
clustering methods and then merges them to reveal the complex
general structure in data. Specifically, their approach reveals a multi-layered group
structure that provides insight into not just the groups but
also their underlying sub-group structure, which may have regular
structure at some level.  We analyze the 13456 HSS of
\citet{misgeldandhilker11} that is briefly summarized in
Section~\ref{data} using syncytial clustering methods 
(introduced in Section~\ref{method}) with
initial clustering results obtained using $t$-mixtures
MBC ($t$MMBC). We analyze the HSS using these methods in
Section~\ref{cl:analysis}. 
Further, in all such investigations involving astronomical
data, parameter estimation in statistical clustering
algorithms is accompanied by uncertainty in those estimates
and careful 	assessment of these estimates is often
required to judge their results 
and their impact on the obtained groupings. We analyze this 
uncertainty by using a nonparametric bootstrap procedure~\citep{efron79}
to calculate the confidence of classification of each data point into
a group. The paper concludes with some discussion about the
physical properties of the obtained groups and to pointers for
future work.


\section{The HSS dataset}
\label{data}
The dataset used in our analysis was compiled from different sources
by~\citet{misgeldandhilker11} and  contains measurements on 13456 HSS of different types: globular clusters (GC), giant
ellipticals (gE), compact ellipticals (cE), ultra compact
dwarf galaxies (UCD), dwarf globular transition objects
(DGTO), nuclear of star clusters (NuSc), bulges of spiral
galaxy (Sbul) and nuclei of nucleated dwarf galaxies (dE,N). There are
12763 HSS in this dataset that are  candidate GCs (GC\_VCC) from the
Virgo Globular Cluster 
survey meaning it is ambiguous if these systems are GC's or
not. \citet{chattopadhyayandkarmakar13} primarily excluded these
candidate GCs from their analysis,
focusing mainly on the 673 stellar systems consisting of confirmed GC's and other different types of stellar systems. We call these 673 systems non-candidate HSS. The main parameters  in \citet{misgeldandhilker11} are
stellar mass ($M_s$), effective radius ($R_e$), mass surface
density averaged over projected effective radius ($S_e$) and
absolute magnitude in the V band ($M_\nu$). Following
\citet{chattopadhyayandkarmakar13} we use the logarithm (base 10) of
$M_s$, $R_e$ and mass-to-luminosity ratio ($M_s / L_\nu$) in
solar luminosity ($M_{s,\odot}$) units in our analysis. The
$M_s/L_\nu$ ratio was obtained using the standard magnitude-luminosity
relation.   
\begin{equation}
	\label{ml:eq}
	\frac{M_s}{M_{s,\odot}} = -2.5\Big(\frac{L_\nu}{L_{\nu,\odot}}\Big)
\end{equation}
where $L_{\nu,\odot}$ denotes the luminosity of the sun. The parameter
$\log_{10}S_e$ is taken into account while interpreting the results
but, as in \citet{chattopadhyayandkarmakar13}, is not used in the clustering mechanism. 
\begin{figure}[h]
	\centering
	\includegraphics[width=\textwidth]{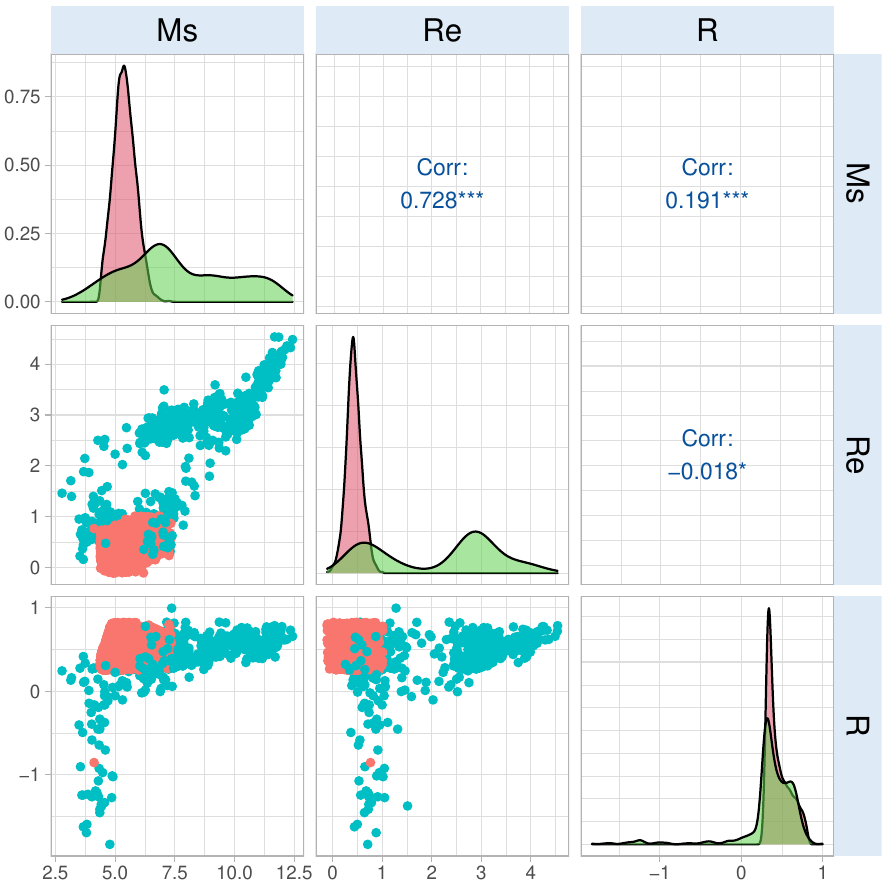}
	\caption{Pairwise scatterplots, estimated densities
		and correlation coefficients of the logarithm (base 10) of
		the		parameters in the HSS dataset. Here Ms denotes mass, Re		effective radius and R the mass-luminosity ratio. In		the scatterplot, orange indicates the non candidates	and blue the candidates. For the density plots the		blue curves represent the non candidates. Note that		the calculated correlations are for all 13456 HSS in		the dataset (including the non-candidates and		candidates).} 
	\label{fig:ggpairs}
\end{figure}
Figure~\ref{fig:ggpairs} provides pairwise scatterplots of the three parameters for both candidate and non-candidate objects along with density plots of the parameters used in clustering.
From Figure~\ref{fig:ggpairs} we see that the densities of the
non-candidates display moderate univariate bimodality, and especially
in the context of effective radius and mass-luminosity ratio. There is
therefore evidence of grouping, certainly so at the univariate level.
Also, the  correlation values indicate that $M_s$ and $R_e$
are  moderately positively-correlated, as expected in stellar objects. But mass-luminosity ratio
has a weak linear association with both  $M_s$ and $R_e$. It therefore
points to complex structure  in the dataset and so  in the coming
sections, we outline methodology to uncover and understand it from
astronomical perspectives.  

\section{Statistical Methodology}
\label{method}
In this section we briefly describe our methodology  for
analyzing the 13456 HSS. Our methodology is a model-based analogue of
the syncytial clustering algorithm of
\citet{riveraandmaitra20}. Our procedure involves an initial
clustering of the data using $t$MMBC followed by merging of the
$t$MMBC mixture components using 
pairwise overlap measure and generalized overlap, calculated using
Monte Carlo methods, as described next.
\subsection{Initial MBC using a mixture of multivariate-\texorpdfstring{$t$} densities}
\label{mbc}
MBC provides a principled way to find homogeneous regular-shaped
groups in a given dataset. It scores
over classical clustering algorithms like $k$-means due to its
ability to better model heterogeneity in groups. As pointed out by
\citet{chattopadhyayandmaitra17} assuming spherically-dispersed homogeneous groups when such assumption is not valid can
lead to erroneous results. In 
MBC~\citep{melnykovandmaitra10,mclachlanandpeel00}
the  observations $X_1,X_2,\ldots,X_N$ are assumed to be
realizations from a \textit{G}-component  mixture
model~\citep{mclachlanandpeel00} with probability density function (PDF)
\begin{equation}
	f(x;\theta) = \sum_{g=1}^{\textit{G}}\pi_g f_g(x; \eta_g)
	\label{mixedmodel}
\end{equation}
where $f_g(\cdot; \eta_g)$ is the density of the $g$th group,
$\eta_g$ the vector of unknown parameters and $\pi_g  = Pr[x_i \in
\mathcal{G}_g]$ is the
mixing proportion of the $g$th group, $g=1,2,\ldots,G$,  and
$\sum_{g=1}^{\textit{G}}\pi_g  = 1$. The component density
$f_g(\cdot; \eta_g)$ can be chosen according to the specific needs of
the application -- the most popular choice being the multivariate
Gaussian density~\citep{chattopadhyayandmaitra17,
	fraleyandraftery98, fraleyandraftery02}. Another useful
family of mixture models proposed by
\citet{mclachlanandpeel98} specifies $f_g(\cdot;
\eta_g)$ to be the multivariate-$t$ density. That is,
\begin{equation}
	\label{mult:t}
	\begin{split}
		f_g(z;
		&
		\mu_g    , \Sigma_g, \nu_g)
		\\ 		&
		= \frac{\Gamma (\nu_g/2 +
			p/2)}{\Gamma(\nu_g/2) {{\nu_g}^{\frac{p}{2}}}{{\pi_g}^{\frac{p}{2}}}{|\Sigma_g|}^\frac{1}{2}}
		\Big[1+ \frac{1}{\nu_g}(z-\mu_g)^T\Sigma_g^{-1}(z-\mu_g)\Big],
	\end{split}
\end{equation}
for $z\in\mathbb{R}^p$, where $\mu_g$ denotes the mean vector,
$\Sigma_g$ the scale matrix and $\nu_g$ the degrees of
freedom, all for the $g$th mixture component, $g=1,2,\ldots,G$.
Our analysis uses, instead of a Gaussian mixture model (GMM),
the multivariate-$t$ mixture model ($t$MM)~\citep{gorenandmaitra22} because the
multivariate-$t$MM better allows for thicker tails in the
component mixture densities.
Subsequent steps in clustering involve obtaining maximum
likelihood (ML) estimates of the parameters       
$\eta_g$, $g = 1, 2,\ldots,G$ using the Expectation-Maximization (EM)
algorithm~\citep{dempsteretal77, mclachlanandkrishnan08,
	chattopadhyayandmaitra17, chattopadhyayandmaitra18} and
assigning each individual observation 
based on the maximum posterior probability that it belongs to
a given group. We use
Bayesian Information Criterion (BIC)~\citep{schwarz78,chattopadhyayandmaitra17} to decide 
$G$. 

\subsection{Merging mixture components using overlap}
\label{merge:full}
After obtaining the parameter estimates of \eqref{mixedmodel},
mixture components can be merged using one of several criteria described in
\citet{melnykov16}. Here we use the pairwise overlap
measure~\citep{maitraandmelnykov10,melnykovandmaitra11,melnykovetal12}. The
pairwise overlap takes value in $[0,1]$ and provides us with a sense of the distinctiveness
between groups obtained using a particular clustering method,
with values closer to 1 indicating that the
two groups are poorly separated. We now provide details on how to
obtain these overlap measures. 
\subsubsection{Pairwise overlap between simple groups}
\label{ol:simple}
A simple group~\citep{riveraandmaitra20} is obtained from the initial
classifications, in our case by $t$MMBC. 
For a $G$ component $t$MM (\ref{mixedmodel}), the probability that an
observation $X$, originally from a $m$th component distribution (that
is, belonging to the $m$th group) gets misclassified to the $n$th group is
\begin{equation}
	\omega_{n|m} = P(\pi_nf_n(x; \mu_n, \Sigma_n, \nu_n) > \pi_mf_m(x; \mu_m, \Sigma_m, \nu_m))
	\label{miss:prob}
\end{equation}
where $f_g(x; \mu_g, \Sigma_g, \nu_g)$ has density as in~\eqref{mult:t}. The pairwise overlap~\citep{maitraandmelnykov10} between the $m$th and $n$th group is then defined as
\begin{equation}
	\omega_{mn} = \omega_{n|m} + \omega_{m|n}
	\label{overlap:form}
\end{equation}
For illustration, we display in the appendix
(Figure~\ref{fig:illus:ol}) sample two-component
three-dimensional distributions with varying degrees of overlap.

Unlike in the case of GMM, the misclassification
probability $\omega_{n|m}$ can not be readily calculated using
analytical methods, so we use Monte Carlo methods as follows:
\begin{enumerate}
	\item Generate $M$ realizations $x_i^*$, $i=1,2,\ldots M$ from
	the density $f_m(x ;\mu_m, \Sigma_m, \nu_m)$.
	\item $\omega_{n|m}$ can then be approximated as
	\begin{equation}
		\begin{split}
			& \hat{\omega}_{n|m}\\
			& = \frac{1}{M}\sum_{i=1}^{M}I\{\pi_nf_n(x_i^*; \mu_n,
			\Sigma_n, \nu_n) > \pi_mf_m(x_i^*; \mu_m, \Sigma_m,
			\nu_m)\} 
		\end{split}
		\label{miss:prob:est}
	\end{equation}
	where $I(\cdot)$ is the indicator function, in this case is 1
	if $\pi_nf_n(x_i^*; \mu_n, \Sigma_n, \nu_n) 
	> \pi_mf_m(x_i^*; \mu_m, \Sigma_m, \nu_m)$ and 0 otherwise.
\end{enumerate}
The pairwise overlap provides us with a sense of the distinctiveness
between groups obtained using a particular clustering method. As
mentioned earlier, these
values lie inside $[0,1]$ with values near zero indicating perfect
separation between groups and those closer to unity indicating  poor
separation. \citet{melnykovandmaitra11} used
methods developed in \citet{maitra10} to define the generalized
overlap $\ddot{\omega}$ to summarize the $G
\times G$ matrix $\Omega$ of pairwise overlaps which is given by
$\lambda_{\Omega}-1/G-1$ where $\lambda_{\Omega}$ is the
largest eigenvalue of $\Omega$. Typically, smaller values of
$\ddot{\omega}$ indicate distinctive groupings.  

\subsubsection{Pairwise overlap between composite groups} 
\label{ol:comp}
Let $\mathcal{G}_p$ and $\mathcal{G}_q$ be two composite groups having
probability densities $\sum_{h \epsilon \mathcal{G}_p }\pi_hf_h(x;
\mu_h, \Sigma_h, \nu_h)$ and $\sum_{r \epsilon \mathcal{G}_q
}\pi_rf_h(x; \mu_r, \Sigma_r, \nu_r)$ respectively, possibly formed by merging one or more components of \eqref{mixedmodel}. The probability of an observation $X$ actually from $\mathcal{G}_p$ being misclassified to $\mathcal{G}_q$ is
\begin{equation}
	\resizebox{0.45\textwidth}{!}{$\omega_{\mathcal{G}_q|\mathcal{G}_p} = P\Bigg(\frac{\sum_{r \epsilon \mathcal{G}_q }\pi_rf_r(x; \mu_r, \Sigma_r, \nu_r)}{\sum_{h \epsilon \mathcal{G}_p }\pi_hf_h(x; \mu_h, \Sigma_h, \nu_h)} > 1\Bigg)$}
	\label{miss:prob:comp}
\end{equation}
where $f_g(x; \mu_g, \Sigma_g, \nu_g)$ has density given by \eqref{mult:t}.
The pairwise overlap between $\mathcal{G}_p$ and $\mathcal{G}_q$,
denoted by $\omega_{\mathcal{G}_p \mathcal{G}_q}$, is then computed as $\omega_{\mathcal{G}_q|\mathcal{G}_p} + \omega_{\mathcal{G}_p|\mathcal{G}_q}$. The misclassification probability $\omega_{\mathcal{G}_q|\mathcal{G}_p}$ can be estimated as follows:
\begin{enumerate}
	\item Generate $M$ samples $x_i^*$, $i=1,2,\ldots M$ from the density $\sum_{h \epsilon \mathcal{G}_p }\pi_h^*f_r(x; \mu_h, \Sigma_h, \nu_h)$,  where $\pi_h^*$ is a standardized probability obtained as $\pi_h^* = \pi_h/\sum_{h \epsilon \mathcal{G}_p } \pi_h$.
	\item$\omega_{\mathcal{G}_q|\mathcal{G}_p}$ can then be approximated as
	\begin{equation}
		\resizebox{0.45\textwidth}{!}{$\hat{\omega}_{\mathcal{G}_q|\mathcal{G}_p} = \frac1M\sum_{i=1}^MI\Big\{\frac{\sum_{r \epsilon \mathcal{G}_q }\pi_rf_r(x_i^*; \mu_r, \Sigma_r, \nu_r)}{\sum_{h \epsilon \mathcal{G}_p}\pi_hf_h(x_i^*; \mu_h, \Sigma_h, \nu_h)} > 1\Big\}$},
		\label{miss:prob:est:comp}
	\end{equation}
	where $I(\cdot)$ is the indicator function, as before. 
\end{enumerate}
\subsection{The model-based syncytial clustering algorithm}
Our syncytial clustering algorithm consists of three phases, the
$t$MMBC phase, the initial overlap calculation phase and the merging
phase: 
\begin{enumerate}
	\item \textit{The initial clustering phase}: This phase fits
	a $G$-component $t$MM to the data using the EM
	algorithm~\citep{mclachlanandpeel00}, with $G$ chosen using
	BIC~\citep{schwarz78}.	
	\item \textit{Merging Phase}: This phase gets triggered only
	if the generalized overlap $\ddot{\omega}^{(1)}$ of the
	original clustering solution is not negligible, that is if
	$\ddot{\omega}^{(1)} \not\approx 0$ and atleast one pairwise
	overlap of the original clustering solution is greater than
	$\ddot{\omega}^{(1)}$. Specifically this phase involves the
	following steps: 
	\begin{enumerate}
		\item Find the pairwise overlaps between $i$th and $j$th group $\omega_{ij}$ for the $G(G-1)/2$ pairs of groups using the method described in Section \ref{ol:simple}. Also find the generalized overlap $\ddot{\omega}$.
		\item Merge the $i$th and $j$th groups if $\omega_{ij}
		> \kappa\ddot{\omega}$, with $\kappa$ as discussed
		shortly,                  and relabel the merged
		groups, as needed.
		\item Calculate the updated pairwise overlaps of the newly formed composite groups using the method described in Section \ref{ol:comp}. Also calculate $\ddot\omega$  of the new clustering solution. Repeat Step (b) with the updated pairwise overlaps and the generalized overlap.
	\end{enumerate}
	\item\textit{Termination Phase}: Terminate the merging in Step 2. if
	the generalized overlap of the current phase clustering
	solution, or its change, is negligible (in this paper defined to be  $\leq 10^{-3}$): that is, if         
	$\ddot{\omega}^{(l)} \approx \ddot{\omega}^{(l-1)}$ or $\ddot{\omega}^{(l)}$   where $\ddot{\omega}^{(l-1)}$ is the generalized overlap of the previous phase clustering solution.
\end{enumerate}
\paragraph*{Selection of $\kappa$}: The parameter $\kappa$ determines the
propensity of merging, and hence the characteristics of the composite
groups formed at each merging phase. With larger values of $\kappa$,
few pairs merge at each phase while smaller values of $\kappa$ mean
that many components are merged simultaneously at each phase. A
data-driven approach proposed by \citet{riveraandmaitra20} for
selecting $\kappa$, that we also adopt, runs the algorithm for several
values of $\kappa$ and uses the final clustering solution with the
smallest terminating $\ddot\omega$.
\paragraph*{Choice of mixture model}: The description of our algorithm
uses $t$MMBC in the initial clustering phase, however, our algorithm
applies to other mixture models with appropriate
modifications. Indeed, in our paper, we have used 
$t$MMBC (which by default, includes GMMBC), with degrees of freedom
also estimated, and then decided on the
initial clustering solution with the higher BIC. 

\subsection{An Illustrative Example}
\begin{figure*}
	\mbox{\subfloat[]{\label{fig:illus:em}\includegraphics[width=0.33\textwidth]{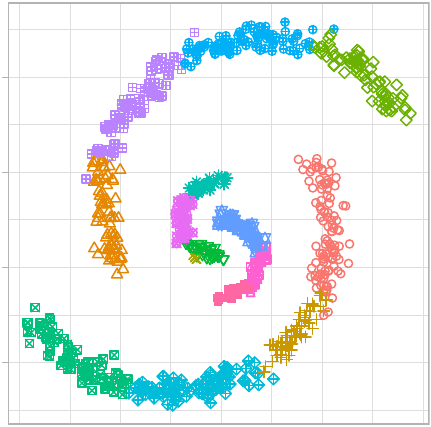}}
		\subfloat[]{\label{fig:illus:ph1}\includegraphics[width=0.33\textwidth]{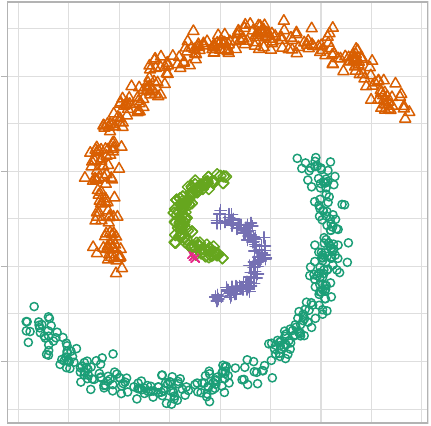}}
		\subfloat[]{\label{fig:illus:ph2}\includegraphics[width=0.33\textwidth]{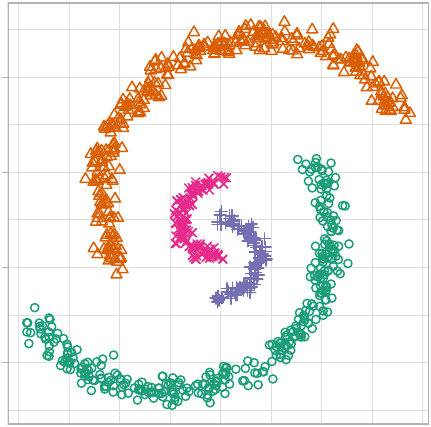}}}
	\caption{From left to right: (a) Original Gaussian mixture model clustering solution according to BIC. (b) Five clusters obtained after first phase merging. (c) Final clustering solution obtained after second phase merging.}
	\label{fig:illus}
\end{figure*} 
We illustrate our model-based syncytial clustering algorithm on the synthetic
2D {\em Bananas-Arcs} dataset in~\citet{riveraandmaitra20}. From
Figure~\ref{fig:illus} it is evident that this dataset has four
complex-structured 
clusters. We fit both GMMBC and $t$MMBC to this dataset using
the methods in~\citet{chattopadhyayandmaitra17}
and~\citet{chattopadhyayandmaitra18} and selected the best clustering
solution as well as the number of groups using BIC. For this dataset,
GMMBC with 15 groups provided the best solution
(Figure~\ref{fig:illus:em}) with $\ddot\omega=8.2\times 10^{-3}$.
The first of the merging phases (with the best $\kappa=3$) provided
the five clusters of Figure~\ref{fig:illus:ph1} with  $\ddot\omega=1.4\times 10^{-4}$. Therefore this merged solution provides a more
distinctive grouping than the Gaussian mixture model solution.  
The second phase of merging resulted in the four groups of 
Figure~\ref{fig:illus:ph2} with 
$\ddot\omega=1.368\times 10^{-4}$. Since the generalized overlap of
this phase is nearly indisinguishable from that obtained at the end of
the first phase of merging, the merging 
terminates here. This example clearly illustrates the
effectiveness of syncytial clustering over regular methods in
capturing complex-structure and irregular-shaped
clusters. Additionally, the multi-staged solution provides us with the
ability to understand each of these complex structures in terms of
their component (simpler) homogeneous groups. Our data-driven approach
results in a terminating $\ddot\omega$ that is much smaller than the
original MBC solution indicating a preference for the
complex-structured solution in providing us with distinctive
groups. We now analyze the 13456 HSS using our syncytial clustering
algorithm. 

\section{Characterization of HSS}
\label{cl:analysis}
\subsection{Cluster analysis}
\subsubsection{The initial clustering phase}
\label{init:clus}
\begin{figure}[h]
	\centering
	\includegraphics[width=\textwidth]{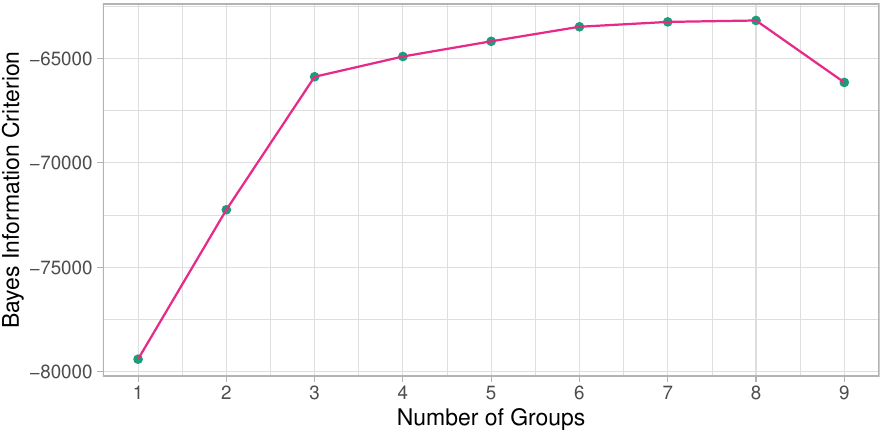}
	\caption{BIC for each $G$ upon performing $G$-component $t$MMBC with 13456 HSS from \citet{misgeldandhilker11}.}
	\label{fig:bic-hss}
\end{figure}

We performed $G$-component $t$MMBC, for $G \in \{1, 2,\ldots,9\}$ on $M_s$, $R_e$ and the
$M_s/L_\nu$, all in the $\log_{10}$-scale, of the 13456 HSS.
Figure \ref{fig:bic-hss} indicates that a eight-component $t$MMBC
provides an optimal fit, as per BIC. The eight-component $t$MMBC solution
was better, as per BIC, than the optimal GMMBC solution. Our results
here are different from that of \citet{chattopadhyayandkarmakar13}
who, upon using $k$-means with the Jump
statistic~\citep{sugarandjames03}, found five groups when excluding
the candidate GCs of \citet{jordanetal08} and four
groups on the full HSS dataset.
\begin{figure}[h] 
	\centering
	\includegraphics[width=\textwidth]{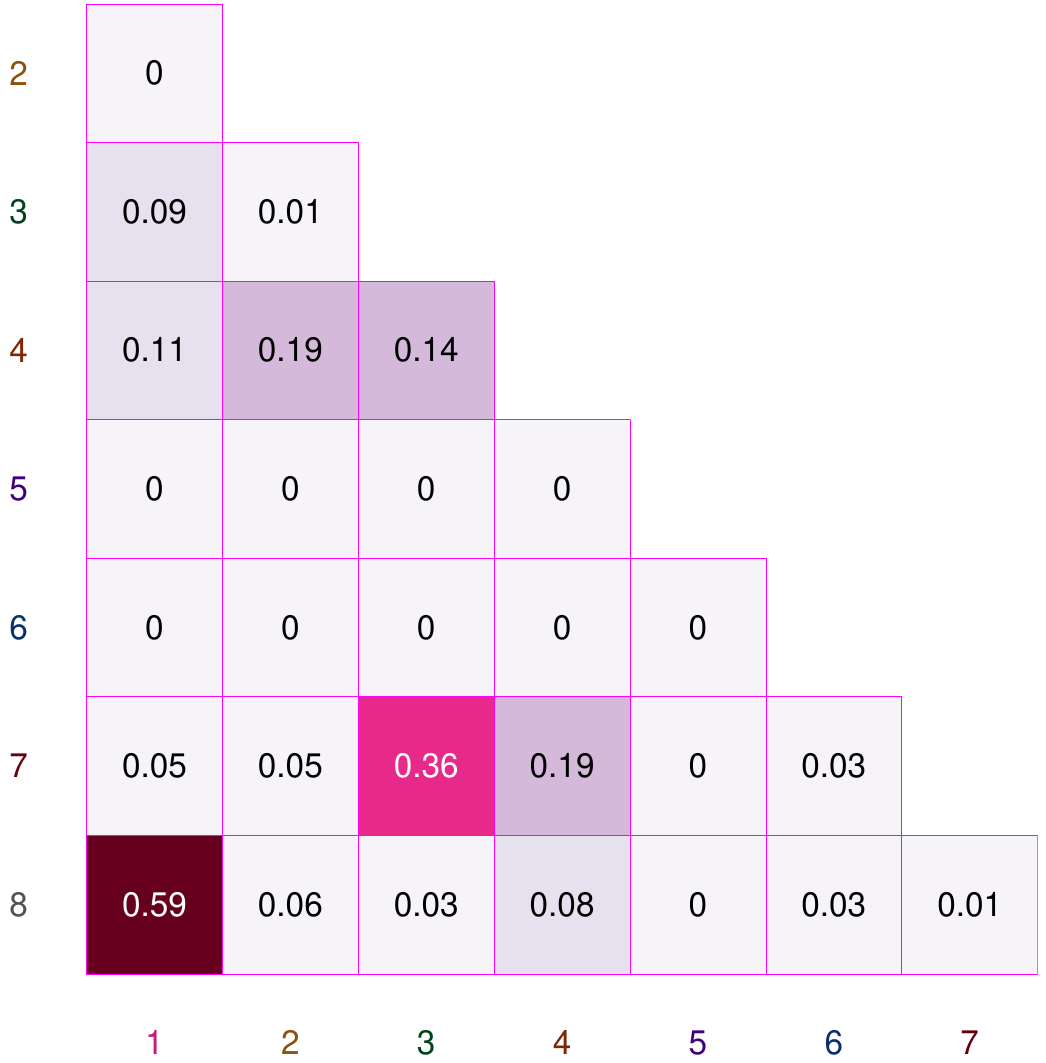}
	\caption{Pairwise overlap measures between any two
		groups obtained by our eight-component $t$MMBC
		solutions.} 
	\label{fig:hss-overlap}
\end{figure}
\begin{figure*}
	\mbox{
		\centering
		\fbox{\subfloat[]{\label{fig:full-pos1}\includegraphics[height=0.45555\textwidth]{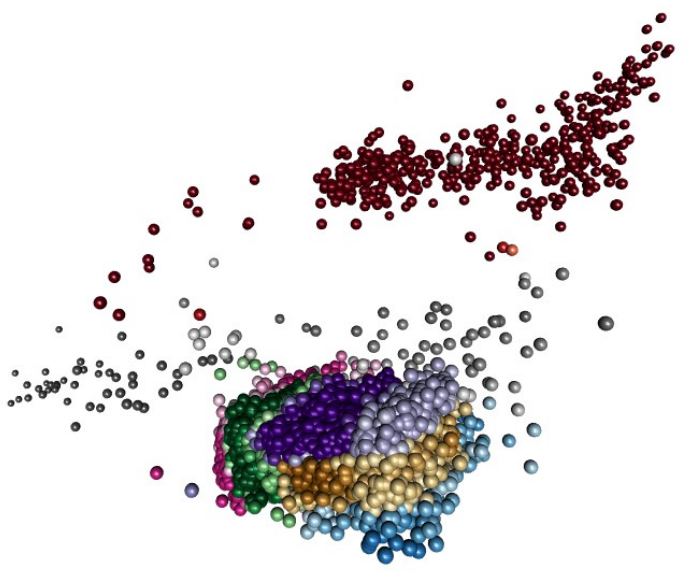}}}
		\fbox{\subfloat[]{\label{fig:full-pos2}\includegraphics[width=0.4325\textwidth]{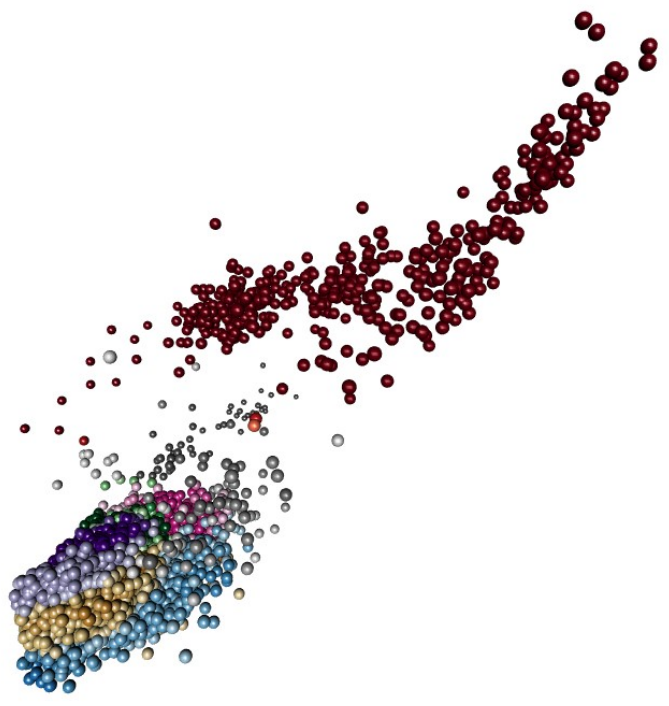}}}
	}
	\mbox{
		\fbox{\subfloat[]{\label{fig:full-pos3}\includegraphics[width=\textwidth]{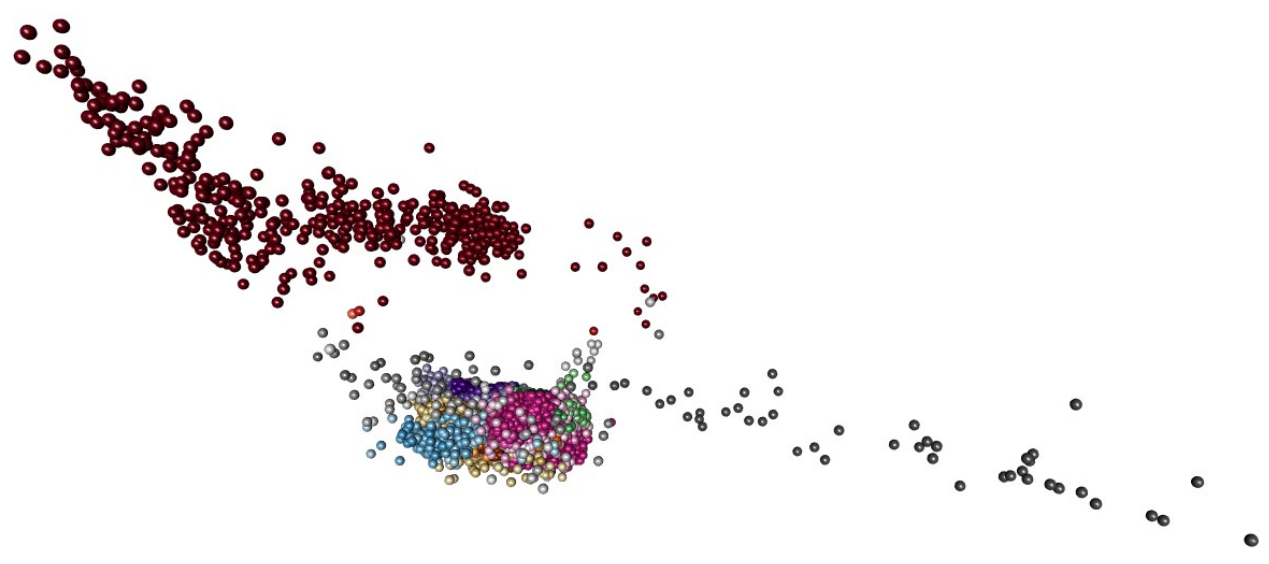}}}
	}
	\caption{Three viewing angles of the scatterplot of the full dataset with different colors representing different groups and intensity of the color signifying the underlying confidence in that particular grouping. Darker shades indicate  higher confidence of classification of that particular object.}
	\label{fig:scatter:all}
\end{figure*}
\begin{table*}
	\caption{Group means and standard deviations (in parentheses)
          of the parameter values for each of the eight groups obtained using $t$MMBC.}
	\label{hss:mean:8}
	\centering
	\begin{tabular}{rrrrrr} \hline\hline
		Group & $M_s$ & $R_e$ & $M_{\nu}$ & $M_\nu/L_\nu$ & $S_e$\\
		\hline
		{\color{Grp1}  1} & 6.184 (0.539)  & 0.331 (0.194) & -9.107 (1.510) & 0.609 (0.119) & 4.723 (0.321)\\
		{\color{Grp2} 2} & 5.134 (0.361) & 0.641 (0.117) & -6.432 (0.844) & 0.630 (0.113) & 3.053 (0.211)\\
		{\color{Grp3} 3} & 5.744 (0.294) & 0.408 (0.057) & -8.609 (0.715) & 0.369 (0.026) & 4.131 (0.3)\\
		{\color{Grp4} 4} & 5.100 (0.326) & 0.470 (0.153) &  -6.905 (0.783) &  0.406 (0.044) & 3.362 (0.387)\\
		{\color{Grp5} 5} & 8.817 (1.939) & 3.032 (0.501) & -16.049 (4.546) &  0.465 (0.153) & 1.954 (1.282)\\
		{\color{Grp6} 6} & 5.803 (1.442) & 0.991 (0.345) &  -9.711 (2.515) & -0.013 (0.707) & 3.023 (1.342)\\
		{\color{Grp7} 7} & 5.313 (0.445) & 0.475 (0.164) &  -7.651 (1.090) &  0.320 (0.022) & 3.565 (0.452)\\
		{\color{Grp8} 8} & 5.556 (0.363) & 0.348 (0.120) &  -7.577 (0.987) &  0.593 (0.098) & 4.062 (0.405)\\ 
		\hline
	\end{tabular}
\end{table*} 
\begin{table*}
	\caption{Types of objects in each of the eight groups obtained by $t$MMBC.}
	\label{hss:or:type}
	\centering
	\begin{tabular}{rrrrrrrrrr} \hline\hline
		& Group {\color{Grp1} 1} & Group {\color{Grp2} 2} &  Group {\color{Grp3} 3} & Group {\color{Grp4} 4} & Group {\color{Grp5} 5} & Group {\color{Grp6} 6} & Group {\color{Grp7} 7} & Group {\color{Grp8} 8}\\ \hline
		cE & 1 &  &  &  & 7 & & & &\\ 
		dE &  &  &  &  & 212 & & & \\
		dE,N & 11 &  & 4  &  &  & 15 & 14 & 1\\
		DGTO &  &   &  &  & 4 &  & 1 & \\
		Dwarf &  &  &  &  & 22 & 2 &  & \\
		GC & 1 & 1 & 5 & 17 & 1 & 58 & 65 & 5\\
		GC\_VCC & 482 & 872 & 1516 & 2814 &  & 10 & 3660  &3409\\
		gE &  &  &  & & 150 & & \\
		NuSc & 6 &  &  &  & 1 & 12 & & 3\\
		Sbul &  &  &  &  & 18  & & \\
		UCD & 14 &  &  &  & 3 & 30 & 8 & 1\\ \hline\hline 
		Total & 515 & 873 & 1525 & 2831 & 418 & 127 & 3748 & 3419\\ \hline
	\end{tabular}
\end{table*}

Figure \ref{fig:hss-overlap} shows the pairwise overlaps of the eight
groups obtained using $t$MMBC. A 3D scatterplot of the eight $t$MMBC
groups is also provided in Figure
\ref{fig:scatter:all}. Table~\ref{hss:mean:8} provides
the group means and standard deviations for the eight groups. The
composition of the eight groups in terms of the different kinds of
stellar objects are in Table~\ref{hss:or:type}. Since $t$MMBC
assumes ellipsoidally-dispersed groups (with fatter tails) some of the
groups exhibit substantial overlap (Figure
\ref{fig:hss-overlap}). These large overlap values point to the
possibility of complex group structure in the data that is not
accounted for by using $t$MMBC (or GMMBC). We therefore explore if we
can use our algorithm to reveal this complex structure.
\subsubsection{The merging phases}
\begin{figure*}
	\mbox{
		\centering
		\fbox{\subfloat[]{\label{fig:mj1-ph1}\includegraphics[height=0.32\textwidth]{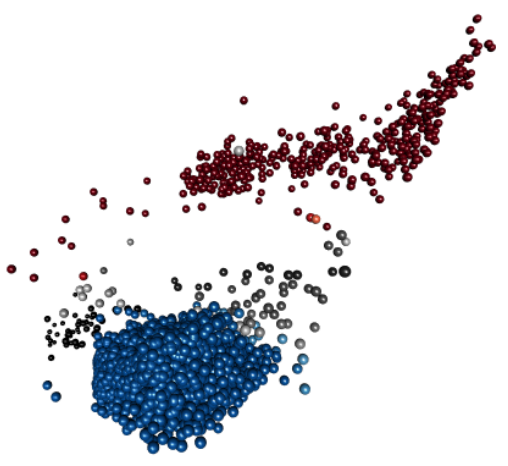}}}
		\fbox{\subfloat[]{\label{fig:mj1-ph2}\includegraphics[width=0.615\textwidth]{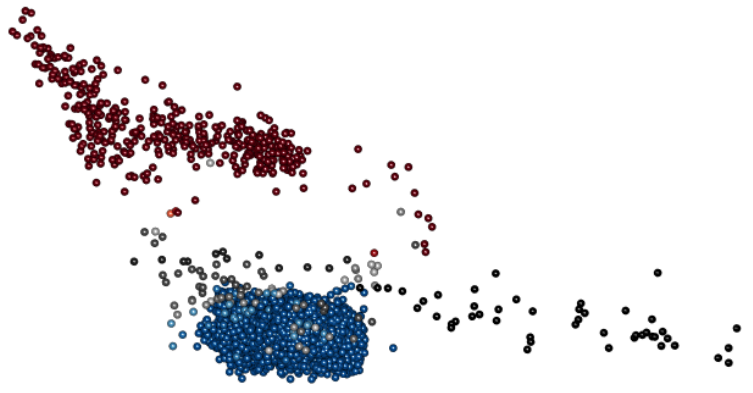}}}
	}
	\caption{Two viewing angles of the scatterplot of the full
		dataset after the first stage merging. Here, different
		colors representing different groups: intensity of the color
		indicatesthe degree of underlying confidence in that particular grouping. Darker shades indicate  higher confidence of classification of that particular object.}
	\label{fig:scatter:ph1}
\end{figure*}

\paragraph{The first phase:}
\label{ph1:res}
Based on the pairwise overlap map of Figure~\ref{fig:hss-overlap}, and
for $\kappa=1$ (determined to be the solution with the lowest
terminating $\ddot\omega$), Groups 1-4, 7 and 8 all merge into one
group. Thus, at the end of this merging phase, we have three groups and a generalized overlap of $\ddot\omega =0.13$. To
facilitate easy reference, we relabel the erstwhile Groups 5 and 6 as
the merged (first phase) Groups (i)-5 and (ii)-6 respectively and the
merged entity as Group (iii), with the component individual groups as
Group (iii)-1 through Group (iii)-4 and Group (iii)-7 and Group
(iii)-8. A 3D scatterplot of the three first-phase merged groups is
provided in Figure~\ref{fig:scatter:ph1}, while
Tables~\ref{hss:mean:ph1} and~\ref{hss:no:type:ph1} summarize the mean
parameter values and the types of objects in the three merged groups.  
The pairwise overlap between
Groups (i) and (ii) is 0.01 and between Groups (ii) and (iii) is
0.1. the overlap between Groups (i) and (iii) is negligible. Figure 
\ref{fig:scatter:ph1} immediately points out how the first phase mergers
based on the pairwise overlaps have been able to capture the
non-ellipsoidal structure present in the data. But the pairwise
overlaps and the generalized overlap (and the logic of our method)
suggest that there might be additional structure which might have
been missed in the first phase merge. So we proceed with another
merging phase. 
\begin{table*}
	\caption{Group means and standard deviations (in parenthesis)
          of parameter values for each of the three groups after the
          first merging phase. The third group is the merged group.}
	\label{hss:mean:ph1}
	\centering
	\begin{tabular}{rrrrrr} \hline\hline
		Group & $M_s$ & $R_e$ & $M_{\nu}$ & $M_\nu/L_\nu$ & $S_e$\\
		\hline
		{\color{Grph1} (i)-5} & 8.817 (1.939) & 3.032 (0.501) & -16.049 (4.546) & 0.465 (0.153) & 1.954 (1.282)\\
		{\color{Grph2} (ii)-6} & 5.803 (1.442) & 0.991 (0.345) & -9.711 (2.515) & -0.013 (0.707) & 3.023 (1.342)\\
		{\color{Grph3} (iii)} & 5.404 (0.467) & 0.438 (0.160) & -7.557 (1.159) & 0.450 (0.139) & 3.731 (0.556)\\
		\hline
	\end{tabular}
\end{table*} 

\begin{table}
	\caption{Types of objects in each of the three groups after phase one merge.}
	\label{hss:no:type:ph1}
	\centering
	\begin{tabular}{rrrrr} \hline\hline
		& Group {\color{Grph1} (i)-5} & Group {\color{Grph2} (ii)-6} &  Group {\color{Grph3} (iii)}\\ \hline
		cE & 6 &  & 1 \\ 
		dE & 212 &  &  \\
		dE,N &  & 15 & 30  \\
		DGTO & 4 &   & 1 \\
		Dwarf & 22 & 2 &  \\
		GC & 2 & 58 & 94 \\
		GC\_VCC &  & 10 & 12753\\
		gE & 150 &  &   \\
		NuSc & 1 & 12 & 9 \\
		Sbul & 18 &  &  \\
		UCD & 3 & 30 & 23 \\ \hline\hline 
		Total & 418 & 127 & 12911\\ \hline
		
	\end{tabular}
\end{table}

\begin{figure*}
	\mbox{
		\centering
		\fbox{\subfloat[]{\label{fig:mj2-ph1}\includegraphics[height=0.39\textwidth]{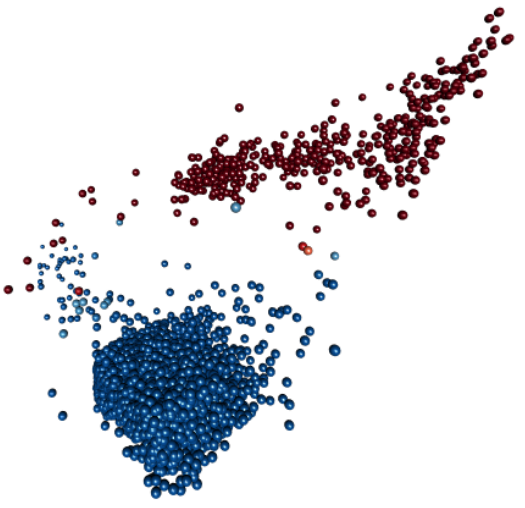}}}
		\fbox{\subfloat[]{\label{fig:mj2-ph2}\includegraphics[width=0.55\textwidth]{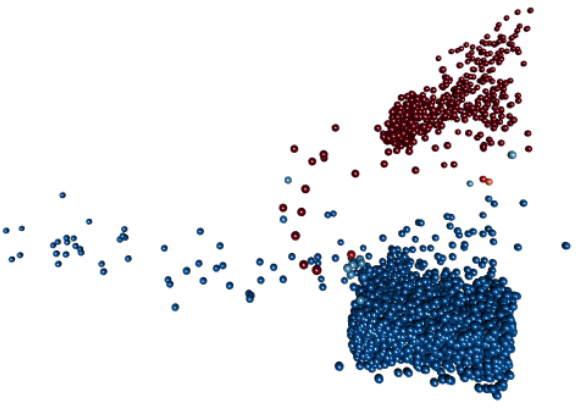}}}
	}
	\caption{Two viewing angles of the scatterplot of the full dataset after third merge with different colors representing different groups and intensity of the color signifying the underlying confidence in that particular grouping. Darker shades indicate  higher confidence of classification of that particular object.}
	\label{fig:scatter:ph2}
\end{figure*}

\paragraph{The second phase:}
In the second phase, Groups (ii)-6 and (iii) merge to yield two
complex-structured groups.  We label the
newly-merged group as 
Group II -- with component merged sub-groups as Groups II-(ii) and
II-(iii), and correspondingly down the hierarchy -- and the original
group as Group I (with sub-groups classification of Group I-(i) and
Group I-(i)-5, to indicate the three-stages of classifications) . A 3D
scatterplot of the  two new groups is given in Figure
\ref{fig:scatter:ph2}, with numerical summaries provided in
Tables~\ref{hss:mean:ph2} and \ref{hss:no:type:ph2}.  The scatterplot
clearly depicts that the 
second merge has captured the complex general-shaped group structures in the
dataset well and also obtained to a good extent the two well-separated groups. 
The pairwise overlap between these two
groups is 0.0013. Since the pairwise overlap between Groups I and II is very close to $10^{-3}$ the algorithm terminates at this stage.
\begin{table*}
	\caption{Group means and standard deviations (in parentheses)
          of the parameters for each of the two groups after the
          second and final  merging phase. The second group is the new merged group.}
	\label{hss:mean:ph2}
	\centering
	\begin{tabular}{rrrrrr} \hline\hline
		Group & $M_s$ & $R_e$  & $M_{\nu}$& $M_\nu/L_\nu$ & $S_e$\\
		\hline
		{\color{Grph1} I} & 8.617 (1.939) & 3.032 (0.501) & -16.049 (4.546) & 0.465 (0.153) & 1.954 (1.282)\\
		{\color{Grph3} II} & 5.408 (0.487) & 0.443 (0.172) & -7.577 (1.199) &  0.445 (0.161) & 3.724 (0.573)\\
		\hline
	\end{tabular}
\end{table*} 

\begin{table}
	\caption{Types of objects in each of the two groups after phase two merge.}
	\label{hss:no:type:ph2}
	\centering
	\begin{tabular}{rrrrr} \hline\hline
		& Group {\color{Grph1} I} & Group {\color{Grph3} II} \\ \hline
		cE & 6 & 1\\ 
		dE & 212 &   \\
		dE,N &  & 45  \\
		DGTO & 4 & 1  \\
		Dwarf & 22 & 2 \\
		GC & 2 & 152 \\
		GC\_VCC & 0 & 12763 \\
		gE & 150 & \\
		NuSc & 1 & 21 \\
		Sbul & 18 &  \\
		UCD & 3 & 53 \\ \hline\hline
		Total & 418 & 13038 \\ \hline
	\end{tabular}
\end{table}

\subsection{Measuring uncertainty in the obtained groupings through bootstrapping}
\label{uncer:hss}
We assess the uncertainty in the parameter estimates in each phase and the
corresponding classification using a nonparametric bootstrap
technique where we  repeatedly resample from the dataset with
replacement and analyze each resample to obtain estimated properties
of the parameters of  interest.  For a detailed study on the
bootstrap refer to~\citet{efron79}.  

In our scenario, the parameters of interest are the group
memberships and a well-done cluster analysis is expected to produce
the same group memberships for most of the observations at each
bootstrap replicate. We now describe the
steps to obtain the confidence of classification of each
observation for the original clustering solution using $t$MMBC.  
\begin{enumerate} 
	\item[(i)] First we perform $t$MMBC on the original
	13456 HSS as described in Section \ref{data}.  This matches the analysis so far and yields 8 $t$MMBC groups. 
	\item[(ii)] Sample, with replacement, 13456 HSS from the
	dataset and again cluster the sampled data using the
	classifications obtained in Step (i) as the initial
	estimates for the groupings. Repeat this procedure $B$ times
	to obtain $B$ sets of classification estimates. Let us
	denote these $B$ sets by $C_1, C_2,\ldots,C_{B}$. Also let
	$C_{ij}$ be 1 if the  classification of the $j$th data point
	in the $i$th bootstrap replicate is the same as the original classification of the $j$th datapoint and 0 otherwise, where $i = 1, 2,\ldots B$ and $j = 1, 2,\ldots,13456$.\item[(iii)] The classification probability for the $j$th data point is obtained as 
	\begin{equation}
		\label{class:prob}
		p_j = \frac{1}{B}\sum_{i=1}^{B} C_{ij}
	\end{equation}
	where $j = 1, 2,\ldots,13456$. In our analysis, we
	took $B=1000$, that is, we resampled 1000 times from
	the dataset.
\end{enumerate}
In each merging phase, the confidence of
classification is obtained by relabeling each of the
$B$ bootstrap replicates to the solution obtained from
the original clustering results and then repeating the steps 
(i)-(iii) with the relabelled bootstrap replicates.
Tables \ref{tab:ph1:conf} and \ref{tab:ph2:conf} give
the number of HSS in each class having confidences in seven
intervals (0, 0.30], (0.30, 0.60], (0.60, 0.85],
(0.85, 0.90], (0.90, 0.95] and (0.95. 1].  The fact
that majority of HSS have a high confidence of
classification in both the stages indicates that we
have been able to achieve good clustering results for
the 13456 HSS.               
\begin{table*}
	\caption{Number of candidate HSS (C) and
		non-candidate HSS (NC) having confidences in each of the seven
		intervals for each of the (a) three groups after the first merge and (b) two groups after the second merge. Here $(\alpha, \beta]$
		denotes the interval with left endpoint $\alpha$ (not included)
		and  right endpoint $\beta$ (included). Entries in the table are left blank when there are no members in that group.}
	\label{tab:conf}
	\mbox{\subfloat[Number of candidates (C) and non-candidates
		(NC) in having confidences  after the first merge.]{\label{tab:ph1:conf}{
				\centering
				\begin{tabular}{ccccccccccccccccc}
					\hline\hline
					Group & \multicolumn{2}{c}{(0, 0.30]} & \multicolumn{2}{c}{(0.30, 0.60]} & \multicolumn{2}{c}{(0.60, 0.80]} & \multicolumn{2}{c}{(0.80, 0.85]} & \multicolumn{2}{c}{(0.85, 0.90]} & \multicolumn{2}{c}{(0.90, 0.95]} & \multicolumn{2}{c}{(0.95, 1]} \\ 
					& NC & C & NC & C & NC & C & NC & C & NC & C & NC & C & NC & C\\\hline
					{\color{Grph1} (i)-5} &  &  &  &  & 1 &  &  &  & 1 &  & 1 &  & 415 & \\
					{\color{Grph2} (ii)-6} &  &  & 8 & 5 & 13 & 2 & 11 & 2 & 28 &  & 12 &  & 46 & 1\\
					{\color{Grph3} (iii)} &  &  & 5 & 4 & 7 & 3 & 1 & 3 & 3 & 3 & 4 & 3 & 138 & 12737\\\hline
				\end{tabular}
			}
		}
	}	
	\mbox{\subfloat[Number of candidates (C) and non-candidates
		(NC) in having confidences after the second merge.]{\label{tab:ph2:conf}{
				\centering
				\begin{tabular}{ccccccccccccccccc}
					\hline\hline
					Group & \multicolumn{2}{c}{(0, 0.30]} & \multicolumn{2}{c}{(0.30, 0.60]} & \multicolumn{2}{c}{(0.60, 0.80]} & \multicolumn{2}{c}{(0.80, 0.85]} & \multicolumn{2}{c}{(0.85, 0.90]} & \multicolumn{2}{c}{(0.90, 0.95]} & \multicolumn{2}{c}{(0.95, 1]} \\ 
					& NC & C & NC & C & NC & C & NC & C & NC & C & NC & C & NC & C\\\hline
					{\color{Grph1} I} &  &  &  &  & 1 &  &  &  & 1 &  & 1 &  & 415 & \\
					{\color{Grph3} II} &  &  & 13 & 9 & 20 & 5 & 12 & 4 & 31 & 3 & 16 & 3 & 184 & 12738\\\hline
				\end{tabular}
			}
		}
	}
\end{table*}

We end our discussion in this section by noting that we
could have also obtained an estimate of the confidence
of classification based on the posterior probability of
classification. However, we use a nonparametric bootstrap 
technique for estimating the classification probabilities in
order to account for the uncertainty in the modeling and to
account for possible model misspecification. The choice of
the nonparametric bootstrap technique provides us
with more robust estimates of the confidence of classification
compared to using the posterior probability of classification because
it accounts for modeling errors to be accounted for in the calculation of
the classification confidence. 

\subsection{Analysis of results}
\subsubsection{Properties of Identified Groups}
Table \ref{hss:or:type} gives the numbers of different types of HSS in each of the $t$MMBC groups and Tables \ref{hss:no:type:ph1}, \ref{hss:no:type:ph2} give the numbers
of different types of HSS in each group at each phase of the
algorithm. The mean and standard deviation of the parameters for the $t$MMBC solution are given in Table \ref{hss:mean:8} while those for each group at each of
the merging phases is in Tables \ref{hss:mean:ph1} and
\ref{hss:mean:ph2}. Additionally, different colors in Figures
\ref{fig:scatter:all}, \ref{fig:scatter:ph1} and \ref{fig:scatter:ph2}
indicate the group to which each HSS belongs, with different shades
indicating the confidence of that particular HSS. Darker shades
indicate higher confidences for that particular datapoint. From the
figures and Tables \ref{tab:ph1:conf} and \ref{tab:ph2:conf} it
is clear that most of the HSS have high confidence coefficients,
indicating 
that the classification method has worked well in identifying the
non-ellipsoidal structure that was demonstrated after the initial
clustering using $t$MMBC. After the first merge, the most notable groups
are Groups (i)-5 and (iii) which have most of the HSS. Most of the
GC\_VCCs are in  Group (iii) after the first merge and only ten are in
Group (ii)-6. The stellar systems in Group (ii)-6 have lower
confidence compared to the other two groups due to which inferences
about the HSS in this group will be more uncertain compared to the HSS in other groups.  From Table \ref{hss:no:type:ph2} we
see that after the second phase, all the candidates (GC\_VCC) were
classified to Group II. All the GCs were also 
classified to this group, as also a number of UCDs and
dE,Ns. These group compositions and  means can help us infer about the
kinematic properties of these stellar systems.  After the final merge,
Group I has a larger $M_s$ and a much larger $R_e$ than Group
II. The effective magnitude is also significantly larger for 
the second group. The $M_s/L_\nu$ ratio is quite similar for
both groups.  We now interpret our results. 
\subsubsection{Interpretations}\label{sec:interpretations}
In this section, we try to understand the physical and evolutionary
properties of the objects in each of the two groups after the final merge using
stellar mass surface density and absolute magnitude. To analyze the
two complex-structured groups at the end of the final merge, we 
carefully look at the properties of the ellipsoidal groups obtained by
$t$MMBC  and group structures obtained with the original eight-groups
clustering solution and after the two mergers, and obtain
deeper insight into the evolutionary and physical properties of these
celestial objects.

From Table \ref{hss:or:type} it is evident that Group 7 is the largest
group obtained by $t$MMBC and also contains the maximum number of
GC\_VCCs . Table \ref{hss:mean:8} indicates that the HSS in these
group have moderate mass and a low effective radius. The surface
density is towards the higher side compared to other groups like Group
5 and this group also has a moderate mass-luminosity ratio. Thus
intuition dictates that the HSS in this group are bright and newer
compared to other groups like Groups 1, 4 and 8. Indeed, the last
group, which is 
mostly composed of GC\_VCCs has a higher mass-luminosity ratio and
surface density compared to Group 7 indicating that that the HSS in
these groups are older compared to that of Group 7 and also
brighter. But the HSS in this group are typically smaller compared to
those of Group 7 and also heavier which gives some indication as to
why the HSS in this group are brighter even though they are older
compared to those of Group 7. Group 4 which is composed of only
globular clusters (both candidate and non-candidate) have properties
similar to those of Group 7. The same can be said about Group 3. This
similarity in terms of kinematic properties also indicates this group
having a high overlap as seen in Figure \ref{fig:hss-overlap}. The same applies for Group 1 which exhibits similarity with Group 8 with respect to kinematic properties. The HSS in Group 2 exhibit kinematic properties that are close to that of Group 4 but they appear to be somewhat smaller than those of Group 4 since they have a higher mass-luminosity ratio compared to that of Group 4. Thus it is very clear that many of the eight $t$MMBC groups exhibit a good degree of similarity in terms of kinematic properties which satisfies the intuition of many of the groups getting merged to reveal only two complex structured groups with a low overlap. We now look proceed to study the kinematic properties of the complex groups obtained at each merging phase. 

At the end of the final merge, Group 1 is essentially Group 5 of the
$t$MMBC solution.  From Table~\ref{hss:mean:ph2}, we see that the HSS
in this group have higher mass and effective radius compared to the
second group. The mass-luminosity ratio for this group is slightly
higher compared to the second group which indicates that the stellar
systems in this group have probably lost most of their massive stars
and are mostly composed of low mass stars, further indicating that
these HSS  are typically older than that of the second
group.  Since this group is similar to Group 5 from the $t$MMBC
solution and did not change during the merging phases so there is no
complex structure present in this group as compared to Group II.  The
conclusions for this group might be a little uncertain for one stellar
object that has a low confidence of classification.  

Group II is much larger in size than  Group I and as mentioned
earlier, contains all the GC\_VCCs and GCs, dE,Ns and most of the
UCDs, with many of the HSS assignments having low
confidence of classification, mainly because this group is formed by
merging Groups (ii)-6 and (iii). So, it would be more helpful to
analyze  this group by separately looking at the properties of Groups
(ii)-6 and (iii) to get a better understanding of the properties of
this group. 

For Group (iii) at the end of the first merging phase, it can be inferred that the
GC\_VCCs in this group (and also Group (ii)-6) are typically smaller
HSS compared to that of Group I (or Group (i)-5). Also the
HSS in this groups have a higher surface density compared
to the other groups formed after the first phase, indicates that these
objects are typically brighter compared to objects in the first group.
These inferences are a little uncertain for the sixteen HSS with low
confidence of classification.  

The stellar systems in Group (ii)-6 have slightly higher mass
compared to Group (iii). The effective radius is also higher for the
HSS in this group which indicates that stellar systems
are larger compared to the objects in Group (iii). The mass-luminosity
ratio is significantly low compared to the other two groups formed in
the first merging phase,and  indicates that these stellar systems are
relatively younger compared to other HSS. A marginally low surface density
compared to that of Group (iii) indicates that these HSS
are not as bright as those in Group (iii) although they
are substantially brighter compared to those of Group (i)-5.
These inferences are highly uncertain since a good number of stellar
systems in this group have low confidence of classification. 

An important aspect of our analysis is how our algorithm is able to
capture the actual number of well-separated groups present in this
data. Previous 
analyses like those of \citet{chattopadhyayandkarmakar13} indicated that there
are four spherically-dispersed groups but the Figures
\ref{fig:scatter:all}, 
\ref{fig:scatter:ph1}, \ref{fig:scatter:ph2} clearly show that the
actual number of well-separated groups is indeed two. The additional
groups found by 
\citet{chattopadhyayandkarmakar13} (and even $t$MMBC in our initial
stage) is due to the inability of $k$-means and $t$MMBC to
detect complex group structures. Our algorithm is able to capture this
complex structure through an effective merging mechanism in multiple
(here, two)
phases. Another important aspect is the confidence of classification
for each HSS that allows us to assess how well our clustering
algorithm performed for this dataset. The fact that most of the HSS
have high confidence of classification indicates that our 
clustering algorithm has been able to reliably capture the complex
structure present in the data. 

\section{Conclusions}
\citet{chattopadhyayandkarmakar13} performed statistical cluster analysis on 673
HSS from \citet{misgeldandhilker11} to determine the homogeneous
groups present in the HSS data. Using the widely-used $k$-means clustering
algorithm along with the jump statistic~\citep{sugarandjames03},
they arrived at five optimal groups whose
properties were explored using fundamental plane relations and
other physical parameters. Their main analysis 
excluded the candidate GCs (GC\_VCCs), because of concerns that
including them would render the data unfit for clustering. Their
analysis hinges on the homogeneous 
spherically-dispersed-groups assumption that underlies
the $k$-means algorithm, which may not be appropriate and can
lead to erroneous results, especially when there is a complex
structure present in the underlying statistical groups as found in
our analysis. We  initially grouped all 13456 HSS of 
\citet{misgeldandhilker11} 
using $t$MMBC and  BIC to optimally find eight groups. Motivated from
a syncytial clustering technique proposed by
\citet{riveraandmaitra20},
we objectively, in a data-driven manner, merged components of the
$t$MMBC solution which revealed two complex-structured groups. Using a
nonparametric bootstrap technique, we further determined the confidence
of classification of each of the 13456 HSS at each stage of merging to
quantify how correctly the stellar systems have been assigned to their
correct groups. We then studied the physical and evolutionary properties of
the objects in each of the two groups by analyzing their mean stellar
mass, surface density and effective radius. 
We found that 
Group I (that is, Group 5 from the original $t$MMBC solution) consists
of typically large and old stellar systems. Group II which is made up
by merging Groups 1-4, 6-8 of the original $t$MMBC solution consists
of younger stellar systems which are brighter than those of Group I
and also smaller in size compared to those in Group I. All the
GC\_VCCs got classified to this group along with all the GCs and some
other HSS. Our  results point to the fact that the GCs are indeed
different from elliptical galaxies which includes the dwarf
ellipticals (dEs). These results support earlier
conclusions~\citep{kormendy85}. Our results also  confirm
that the GC\_VCCs are most likely to be GCs and that there is very less
chance that these are dwarfs.

A reviewer has asked whether our merging stages can, for instance,
result in erasing of correlations between parameters that may have
physical meaning, and whether this presents a limitation for our
method. We  note that correlation only measures linear
association (or relationships)  and is inadequate for describing other
kinds of relationships. The possible reduction or erasure of the
correlation between parameters in complex-structured groups formed by
merging, is a consequence more of the complex structure of the well-separated
groups: note also that the relationships (including linear 
ones) continue to be described, as necessary, at the appropriate subgroup
level. Therefore, we do not necessarily consider our methodology to
have a limitation here: rather, we feel that it provides a more
nuanced understanding 
of the group structure present in the HSS data. Finally, we note that
the goal of clustering is to find different groups of homogeneous
observations. How the clustering is done is 
dependent on the resolution (which is analogous to the merging level)
and the properties that we would like our clusters to portray. For
instance, if we want  ellipsoidally-shaped groups, then the
eight-component $t$MMBC
solution provides a good description of the HSS. At the same time, we
can see which of these groups of HSS are closer (or more similar) to
each other, at the different resolution levels, and how they form into
more generally-shaped groups. Our approach here formally helps
determine the groups that combine ({\em i.e.}, are similar) at each 
level and an objective approch to describe where the groups are
distinct enough to require separate consideration and understanding of
their properties. As we see in Section~\ref{sec:interpretations}, this
sort of understanding of the underlying complex group structure can
better inform our understanding of the characteristics of the
different kinds of HSS. 

There are a number of issues that merit further
investigation. For example it would be useful to explore if
the logarithmic transformation used on the parameters is
plausible or is detrimental to the objective of finding groups of
HSS. It is possible that some other 
transformation may lead to better-separated
groups. Our analysis may also be extended to larger and more
comprehensive data sets than in this study, in order to analyze
whether similar results hold for the HSS from other sources. Also
with additional information like temperature and
color indices of the stars present in the system, it might be
possible to obtain  deeper understanding of the evolution of
these systems possibly through HR diagrams. Apart from this it would
also be interesting to analyze how fundamental plane relations hold in
conjunction to these results. Thus we see that while our analysis has
provided some interesting insights into 
the different types of HSS, additional issues remain that
deserve further attention. 
\begin{acknowledgement}
We sincerely thank T. Chattopadhyay for providing us with the dataset
of \citet{chattopadhyayandkarmakar13}. We are also very grateful to
C. Struck for helpful comments on clarifying and 	interpreting
our results. Our sincere appreciation also to an Associate Editor and
an anonymous reviewer 
	whose thorough and insightful comments greatly improved an earlier
	version of this article.
\end{acknowledgement}
\appendix
\section{Illustration of the pairwise overlap}	
\label{appendix}
\begin{figure*}[!h]
	\mbox{\fbox{\subfloat[]{\label{fig:illus:ol1}\includegraphics[width=0.27\textwidth]{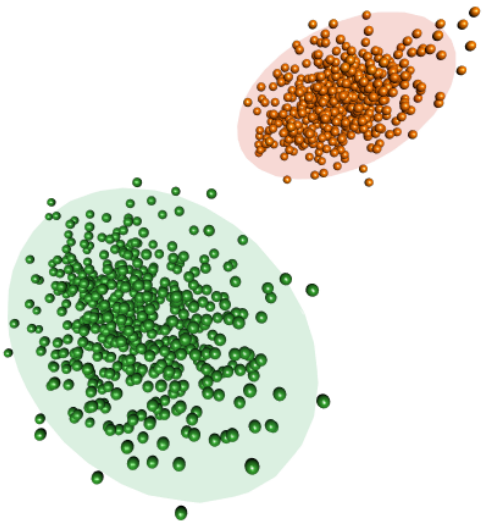}}}
		\fbox{\subfloat[]{\label{fig:illus:ol2}\includegraphics[width=0.435\textwidth]{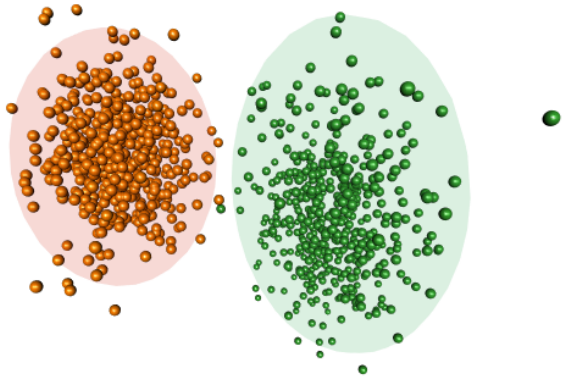}}}
	\mbox{\fbox{\subfloat[]{\label{fig:illus:ol3}\includegraphics[height=0.22\textheight]{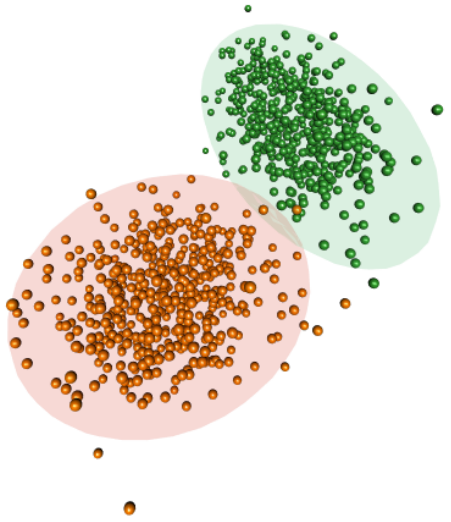}}}}}
	\mbox{\fbox{\subfloat[]{\label{fig:illus:ol4}\includegraphics[height=0.302\textwidth]{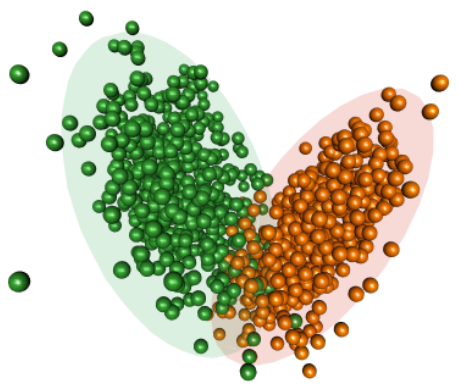}}}
	\fbox{\subfloat[]{\label{fig:illus:ol5}\includegraphics[width=0.317\textwidth]{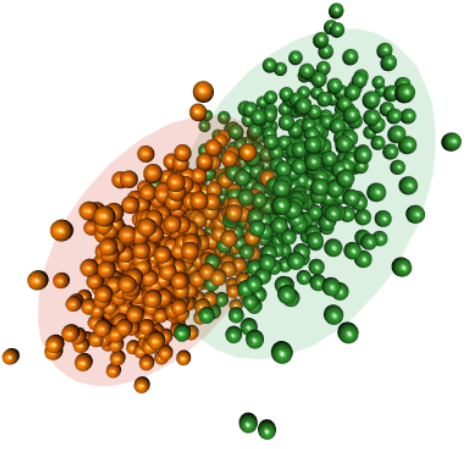}}}
	\fbox{\subfloat[]{\label{fig:illus:ol6}\includegraphics[height=0.302\textwidth]{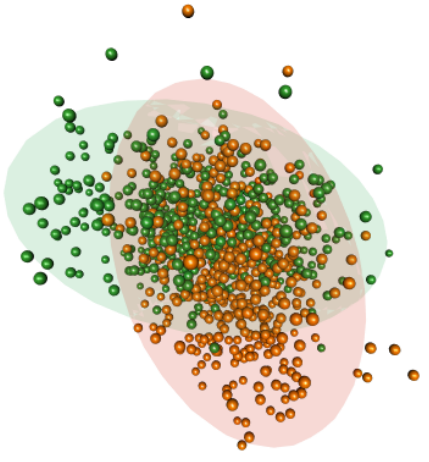}}}}
      \caption{Realizations from sample two-component Gaussian mixture
        distributions having pairwise overlap measures of (a) $\omega
        = 0.00001$, (b) $\omega = 0.001$, (c) $\omega = 0.01$
        (d) $\omega =  0.05$, (e) $\omega = 0.1$ and (f) $\omega = 0.5$. For each
        component, we also provide the 95\% ellipsoid of concentrations.}
	\label{fig:illus:ol}
\end{figure*} 
In this section, we demonstrate the kinds of 3D distributions that may be
obtained with specification of different pairwise overlap measures.
Using the R~\citep{R} package {\sc MixSim} \citep{melnykovetal12}, we
simulated 1000 observations from two-component GMMs with pairwise
overlap $\omega =  10^{-5}, 0.001,\allowbreak 0.01, 0.05, 0.1$, and $0.5$. The simulated  data are
displayed in Figure \ref{fig:illus:ol}. Each component distribution is
also specified by  means of its 95\% ellipsoid of concentration. The
figure shows very good separation with $\omega=10^{-5}$, and even with
$\omega=0.001$, modest separation for  $\omega=0.01$, 
substantially poor separation for $\omega=0.05$, and worsening
separation with higher values of $\omega$.


\bibliography{references}

\appendix

\end{document}